\documentclass[a4paper]{spie}  

\usepackage{amsmath,amsfonts,amssymb}
\usepackage{graphicx}
\usepackage[colorlinks=true, allcolors=blue]{hyperref}

\usepackage{soul}
\usepackage{siunitx}
\usepackage{glossaries}
\usepackage{cleveref}
\usepackage{rotating}

\title{Design of the ULTRASAT UV Camera} 

\author[1]{Arooj Asif}
\author[1]{Merlin Barschke}
\author[1]{Benjamin Bastian-Querner}
\author[1,2]{David Berge}
\author[1]{Rolf B\"uhler}
\author[1]{Nicola De Simone}
\author[1]{Gianluca Giavitto}
\author[1]{Juan M. Haces Crespo}
\author[1]{Nirmal Kaipachery}
\author[1,2]{Marek Kowalski}
\author[1]{Shrinivasrao R. Kulkarni}
\author[4]{Daniel K\"usters}
\author[1]{Sebastian Philipp}
\author[1]{Heike Prokoph}
\author[1,2]{Julian Schliwinski}
\author[1]{Mikhail Vasilev}
\author[1]{Jason J. Watson}
\author[1]{Steven Worm}
\author[1]{Francesco Zappon}
\author[5]{Shay Alfassi}
\author[3]{Sagi Ben-Ami}
\author[5]{Adi Birman}
\author[6]{Kasey Boggs}
\author[6]{Greg Bredthauer}
\author[5]{Amos Fenigstein}
\author[3]{Avishay Gel-Yam}
\author[5]{Dmitri Ivanov}
\author[5]{Omer Katz}
\author[3]{Ofer Lapid}
\author[3]{Tuvia Liran}
\author[3]{Ehud Netzer}
\author[3]{Eran O. Ofek}
\author[7]{Shirly Regev}
\author[3]{Yossi Shvartzvald}
\author[6]{Joseph Tufts}
\author[5]{Dmitry Veinger}
\author[3]{Eli Waxman}
\affil[1]{Deutsches Elektronen-Synchrotron, Platanenallee 6,  D-15735 Zeuthen, Germany}
\affil[2]{Institut f\"ur Physik, Humboldt-Universit\"at zu Berlin, Newtonstrasse 15, D-12489 Berlin, Germany}
\affil[3]{Department of Particle Physics and Astrophysics, Weizmann Institute of Science, Herzl St 234, Rehovot, Israel}
\affil[4]{Department of Physics, University of California at Berkeley, 366 LeConte Hall MC 7300, Berkeley, CA, 94720-7300}
\affil[5]{Tower Semiconductor, 20 Shaul Amor Avenue, Migdal Haemek, 2310502, Israel}
\affil[6]{Semiconductor Technology Associates, Inc., 1241 Puerta Del Sol, San Clemente, CA 92673-6310, USA}
\affil[7]{Etesian Semiconductor Ltd, P.O.B 3227, Ramat Yishai, 3004205, Israel}

\authorinfo{Send correspondence to:\\$^*$Rolf Bühler: E-mail: rolf.buehler@desy.de\\  $^\dagger$Merlin Barschke: E-mail: merlin.barschke@desy.de}

\newacronym{DA}{DA}{Detector Assembly}
\newacronym{RE}{RE}{Remote Electronics}
\newacronym{FME}{FME}{Focus Mechanism}
\newacronym{NUV}{NUV}{Near Ultraviolet}
\newacronym{CMOS}{CMOS}{Complementary metal-oxide-semiconductor}
\newacronym{QE}{QE}{Quantum Efficiency}
\newacronym{CAD}{CAD}{Computer Aided Design}
\newacronym{STA}{STA}{Semiconductor Technology Associates}
\newacronym{PDR}{PDR}{Preliminary Design Review}
\newacronym{TJ}{TJ}{Tower Semiconductor Ltd.}
\newacronym{AV}{AV}{Analog Value Ltd.}
\newacronym{GALEX}{GALEX}{Galaxy Evolution Explorer}
\newacronym{WIS}{WIS}{Weizmann Institute of Science}
\newacronym{DESY}{DESY}{Deutsches Elektronen Synchrotron}
\newacronym{ULTRASAT}{ULTRASAT}{Ultraviolet Transient Astronomical Satellite}
\newacronym{BSI}{BSI}{Back-Side Illuminated}
\newacronym{ARC}{ARC}{Anti-Reflective Coating}
\newacronym{REB}{REB}{Remote Electronics Board}
\newacronym{SBB}{SBB}{Sensor Bias Board}
\newacronym{LDO}{LDO}{Low-Dropout Regulator} 
\newacronym{FPGA}{FPGA}{Field Programmable Gate Array} 
\newacronym{ISA}{ISA}{Israel Space Agency}
\newacronym{SPI}{SPI}{Serial Peripheral Interface}
\newacronym{LVDS}{LVDS}{Low-Voltage Differential Signaling}
\newacronym{PCB}{PCB}{Printed Circuit Board}

\begin{document} 
\maketitle

\begin{abstract}
The \gls{ULTRASAT} is a scientific UV space telescope that will operate in geostationary orbit. The mission, targeted to launch in 2024, is led by the \gls{WIS} in Israel and the \gls{ISA}. \gls{DESY} in Germany is tasked with the development of the UV-sensitive camera at the heart of the telescope. The camera’s total sensitive area of $\approx$~\SI{90x90}{\milli\meter} is built up by four back-side illuminated CMOS sensors, which image a field of view of $\approx$~\SI{200}{deg\squared}. Each sensor has \num{22.4} megapixels. The Schmidt design of the telescope locates the detector inside the optical path, limiting the overall size of the assembly. As a result, the readout electronics is located in a remote unit outside the telescope. The short focal length of the telescope requires an accurate positioning of the sensors within  \SI{\pm50}{\micro\meter} along the optical axis, with a flatness of \SI{\pm10}{\micro\meter}. While the telescope will be at around \SI{295}{\kelvin} during operations, the sensors are required to be cooled to \SI{200}{\kelvin} for dark current reduction. At the same time, the ability to heat the sensors to \SI{343}{\kelvin} is required for decontamination. In this paper, we present the preliminary design of the UV sensitive \gls{ULTRASAT} camera. 
\end{abstract}


\section{INTRODUCTION}
\label{sec:intro}  

\gls{ULTRASAT} is an astronomical satellite mission whose wide-angle telescope ($\approx$200 deg$^2$) will image the sky in the \gls{NUV} between 220~nm and 280~nm. The mission is designed to survey large fractions of the sky at high cadence (minutes to months) and sensitivity (AB limiting magnitude 22.3 AB at 5$\sigma$ in 3$\times$\SI{300}{\sec} integration). The volume of the Universe that will be surveyed per unit time by \gls{ULTRASAT} is more than 300 times larger than the one accessible to the most sensitive UV satellite flown to date, the \gls{GALEX}\cite{2007ApJS..173..682M}. For hot sources with temperatures exceeding \SI{20000}{\kelvin} the volumetric survey speed will be comparable to upcoming surveys as the Vera C. Rubin Observatory \cite{2014AJ....147...79S}. The main science objectives of \gls{ULTRASAT} are the detection of counterparts to gravitational wave sources and of supernovae \cite{2016ApJ...820...57G,2018ApJ...855L..23A}. Other science topics are variable and flaring stars, active galactic nuclei, tidal disruption events, compact objects, and galaxies. The satellite is expected to be in orbit by the end of 2024 and will take data for at least three years.  \gls{ULTRASAT} is jointly funded and managed by the \gls{ISA} and the \gls{WIS}, under the scientific leadership of the \gls{WIS}. \gls{DESY} in Germany joined the mission in 2019 and will provide the camera (i.e. the Focal Plane Array and the readout electronics).

The \gls{ULTRASAT} camera design addresses several technical challenges at the same time. First, a high photon sensitivity needs to be achieved in the \gls{NUV} over a wide sensitive area. A mosaic of four \gls{CMOS} sensors with 22.4 megapixels and a sensitive area of $\approx 4.5 \times 4.5$~cm$^2$ each was designed for this purpose. Second, the sensor needs to be operated at \SI{200}{\kelvin} to achieve the low noise level required for deep observations. At the same time heating the sensor to \SI{343}{\kelvin} needs to be possible for decontamination. Third, a flatness of \SI{\pm10}{\micro\meter} over the entire focal plane is required for the fast optics of the telescope (f/$\approx 1.1$). Finally, the camera needs to be as compact as possible to limit obscuration, as it is located in the light path inside the telescope.

This paper describes the camera design at \gls{PDR} in December 2020. A outlook of changes made since then will be given at the end of this publication. The article is structured as follows: in section~\ref{sec:design}, we give an overview of the camera design. In section~\ref{sec:sensor}, we will describe the main sensor parameters and the mosaic assembly. We proceed to describe the mechanical and thermal design of the camera in section~\ref{sec:structure}. Afterwards, the readout electronics is described in section~\ref{sec:electronics}. Finally, we conclude introducing some design updates that were done after \gls{PDR} in section~\ref{sec:conclusions}.

\section{Camera design overview}

\begin{figure} [bt]
   \begin{center} 
   \includegraphics[width=1\textwidth]{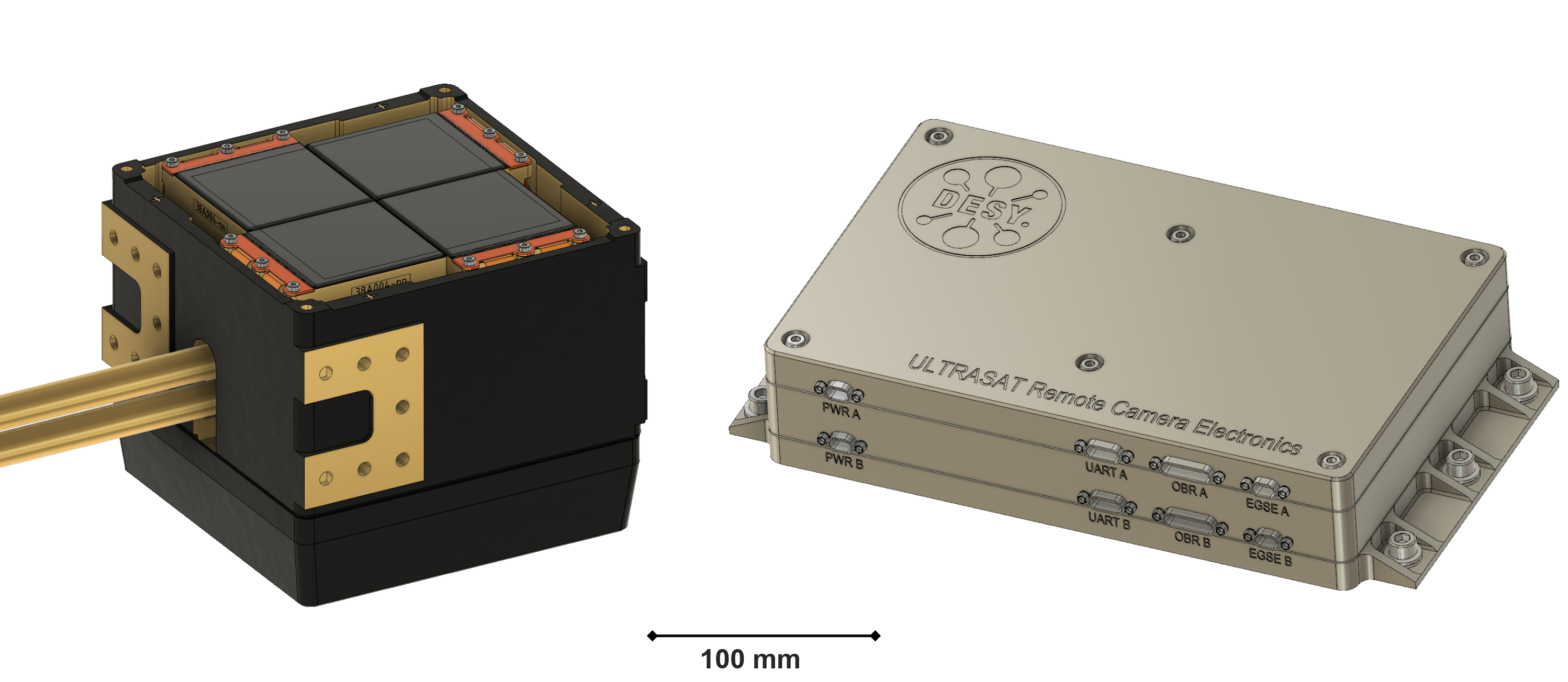}
   \caption[] 
   { \label{fig:cam_dim} 
Digital renderings of the Detector Assembly (left) and the Remote Electronics (right) of the \gls{ULTRASAT} camera.}
 \end{center}
   \end{figure}

The camera is composed of two main elements: the \gls{DA} and the \gls{RE}, shown in Figure \ref{fig:cam_dim}. The \glspl{REB} within the \gls{RE} control and read out the sensors and communicate with the spacecraft. The \glspl{SBB} within the DA perform power conditioning to supply the sensors. The sensor is cooled by means of thermal straps in combination with heat pipes. The fine control of the sensors’ temperature as well as the heating of the sensors for decontamination is performed by the spacecraft by means of a control loop using temperature sensors and heaters installed in the \gls{DA}. 

\label{sec:design}

The power and data interfaces between spacecraft and camera are located at the \gls{RE} unit outside the telescope. The data interface can be divided between a command and telemetry interface, as well as a high speed interface for downloading the image data to the spacecraft. Mechanically, the \gls{DA} is mounted to the telescope, whereas the \gls{RE} is mounted to the spacecraft’s structure. The outer dimensions of the \gls{DA} assembly excluding the heat pipes, the cabling and externally attached elements are $\approx$~\SI{135x135x120}{\milli\meter}. Its total mass is expected to be  $\approx$~\SI{5.6}{\kilo\gram}. The \gls{RE} has dimensions of $\approx$~\SI{264x153x49}{\milli\meter}. Its total mass is expected to be $\approx$~\SI{2.5}{\kilo\gram}.

\begin{figure} [tb]
   \begin{center}
   \includegraphics[width=0.75\textwidth]{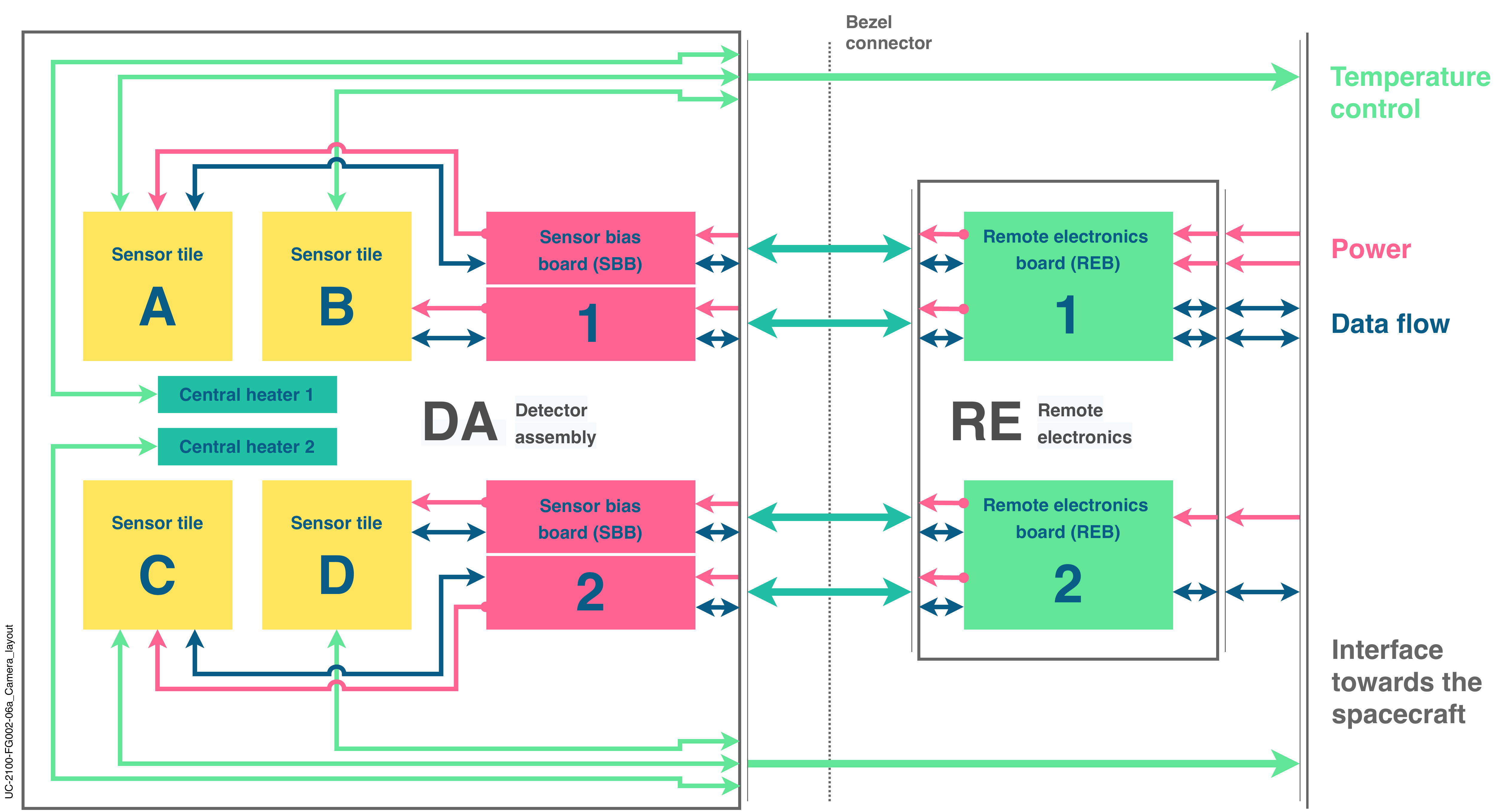}
   \end{center}
   \caption[] 
   { \label{fig:cameraOverview} 
Schematic drawing of the data and power flow in the \gls{ULTRASAT} camera.}
   \end{figure}

The overall electrical layout of \gls{ULTRASAT} camera is shown in Figure~\ref{fig:cameraOverview}. The four identical sensor tiles are assembled on a mosaic plate. Each two tiles are connected to one \gls{SBB} which carries Point of Loads required to bias the sensor. Each sensor tile carries temperature sensors and heaters that allow for fine-controlling its temperature. These heaters and temperature sensors are directly connected to the spacecraft. A total of four cables connect the \gls{DA} to the \gls{RE}. The two \glspl{REB} inside the \gls{RE} hold the \glspl{FPGA} to read out the sensors and the main power converters supplied by the spacecraft’s unregulated power lines. The camera design follows a graceful degradation  approach. This means, that a failure of a component may lead to degraded camera performance, but not to a global system failure. For instance, if one of the four sensors fail, the others still remain operational.

\section{Sensor and mosaic assembly}
\label{sec:sensor}

The \gls{CMOS} sensors were designed and produced for the \gls{ULTRASAT} mission by \gls{TJ} and \gls{AV} in Israel. The sensor's main performance parameters are shown in Table~\ref{tab:sensor}. A key performance goal was to achieve a high \gls{QE} in the \gls{NUV}. For this purpose, the sensors are back-illuminated and coated with an \gls{ARC}. The different \glspl{ARC} were optimized for the \gls{NUV} by \gls{TJ}. Their performance was tested by \gls{DESY} with test sensors provided by \gls{TJ}. The chosen coating has a high average \gls{QE} of $\approx$~\SI{60}{\percent} in the range of \SIrange[range-units=single]{220}{280}{\nano\meter}. Note, that this value is corrected for quantum yield which is significant at these frequencies. The measurement setup and the results are discussed in more detail in an accompanying article \cite{SPIEScouts, 2020SPIE11447E..71K}. The detailed sensor design and the ongoing sensor performance validation will be published in forthcoming publications.

\begin{table}[tb]
\caption{\label{tab:sensor} Main design specifications for one \gls{CMOS} sensor of the \gls{ULTRASAT} camera mosaic.} 
\begin{center}       
\begin{tabular}{|l|l|} 
\hline
\rule[-1ex]{0pt}{3.5ex}  \textbf{Characteristic} & \textbf{Design specification}  \\
\hline
\rule[-1ex]{0pt}{3.5ex}  Pixel size & 9.5 $\mu$m $\times$ 9.5 $\mu$m   \\
\hline
\rule[-1ex]{0pt}{3.5ex}  Photo sensitive area & 45.011 mm $\times$ 45.011 mm (4738 $\times$ 4738 pixels)  \\
\hline
\rule[-1ex]{0pt}{3.5ex}  Die size & 47.135 mm $\times$ 50.080 mm  \\
\hline
\rule[-1ex]{0pt}{3.5ex} Mean \gls{QE} between 220-280nm  & $>$~60$\%$  \\
\hline 
\rule[-1ex]{0pt}{3.5ex}  Low-gain full well capacity &  $>$~140ke- \\
\hline 
\rule[-1ex]{0pt}{3.5ex}  Dark current at 200 K & $<$~0.026 e-$/$s  \\
\hline 
\rule[-1ex]{0pt}{3.5ex}  Readout noise & $<$~3.5 e-   \\
\hline 
\rule[-1ex]{0pt}{3.5ex}  Readout mode &  Rolling shutter \\ 
\hline 
\rule[-1ex]{0pt}{3.5ex}  Readout time &  $<$~20 sec \\ 
\hline 
\end{tabular}
\end{center}
\end{table}

\begin{figure} [tb]
   \begin{center}
   \includegraphics[width=0.75\textwidth]{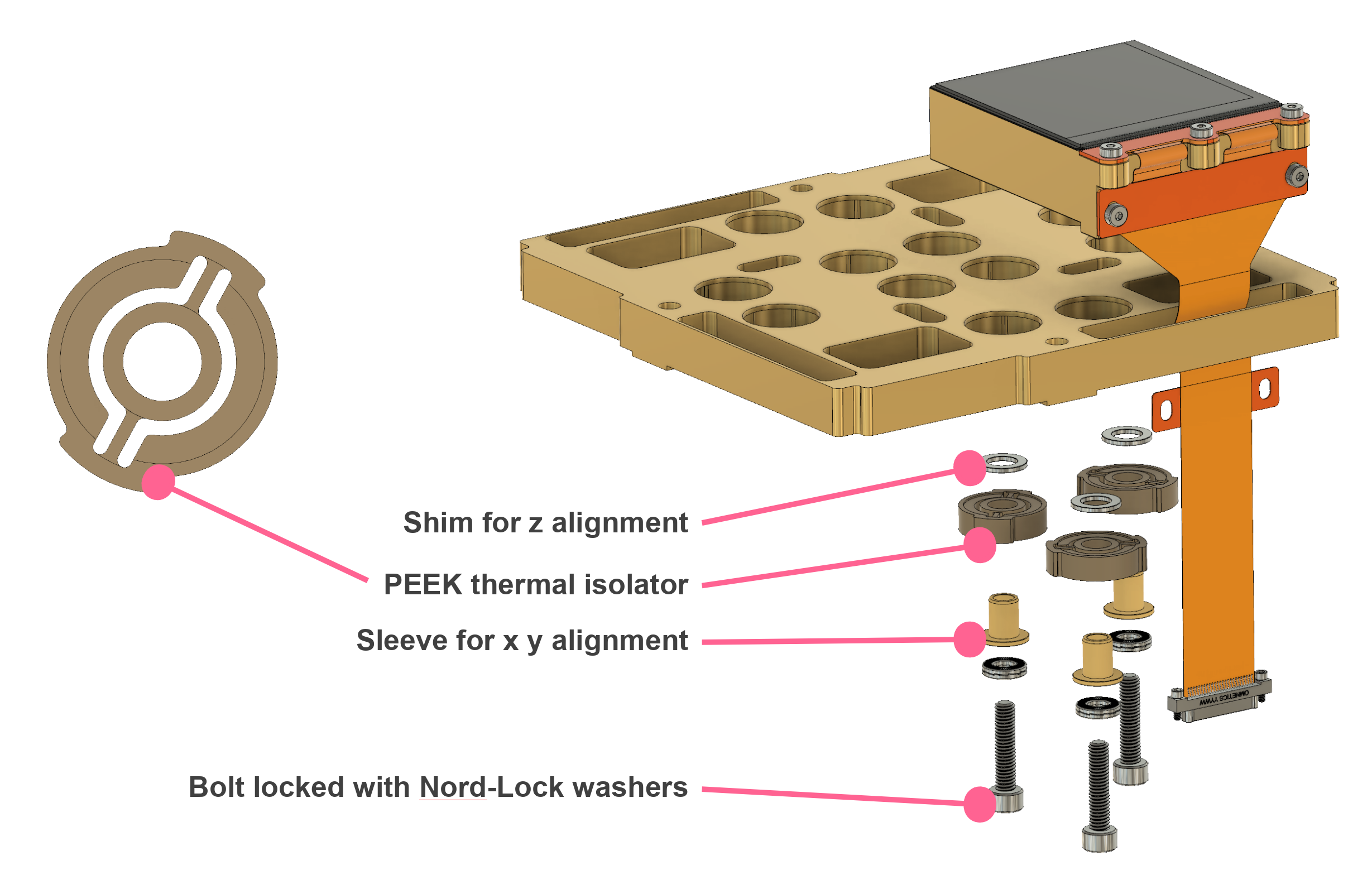}
   \end{center}
   \caption[] 
   { \label{fig:STAPDR} 
\gls{CAD} drawing of the sensor package and mosaic assembly. The package and mosaic assembly was designed by \gls{STA} and shows the design at the \gls{PDR} of the camera.}
   \end{figure}

The mosaic assembly was designed by the company \gls{STA} in California. The assembly consists of a sensor package, a mosaic plate and a sensor flex \gls{PCB} for noise filtering shown in Figure~\ref{fig:STAPDR}. Four \gls{CMOS} sensor tiles are aligned in a windmill configuration. This configuration has two main advantages: first, it allows to minimize the insensitive area in between sensors to $\lessapprox$~\SI{2}{\milli\meter}, as the \gls{CMOS} periphery electronic is always at the edge of the mosaic. Second, all sensor tiles are interchangeable, making it possible to select the best sensor tiles for the flight camera.

The sensors are glued to the package with Epotek 301-2 epoxy. The sensor package will be fabricated from CE6, a silicon-aluminum composite from Sandvik-Osprey, to match the expansion coefficient of the sensors. Heli-Coils are installed in the package for mounting the flex circuit as well as the mosaic plate. The flex circuit is wire bonded to the sensor using aluminum wedge wire bonding. The wire bonds are the highest protrusion from the assembly, $<$200~$\mu$m above the sensor. Z shims are adjusted on each of three feet per package to make the four tiles in a mosaic co-planar. A high flatness significantly below 20~$\mu$m will be achieved by adjusting the shim thicknesses. There are three thermal isolators per package, made from PEEK material. The isolators are placed such that the flexible dimension is oriented radially to allow compliant shrinkage of the tile relative to the warm temperature stable mosaic plate. Note, that the design described here has evolved further since \gls{PDR}, as will be discussed in section~\ref{sec:conclusions}.

\section{Mechanical and thermal design}
\label{sec:structure}

The thermo-mechanical design of the \gls{DA} is mainly determined by two requirements: first, the sensors are required to be positioned with an absolute accuracy of \SI{<50}{\micro\meter} and a flatness of \SI{\pm10}{\micro\meter} along the optical axis. Second, the temperature of the sensor needs to be between \SI{195}{\kelvin} and \SI{205}~K during observations and at $\approx$\SI{ 343}{\kelvin} for in-orbit decontamination. At operation the sensor's temperature  needs to be homogeneous and stable to \SI{1}{\kelvin} within the nominal observing time of 300 sec.

\begin{figure} [tb]
   \begin{center}
   \includegraphics[width=0.95\textwidth]{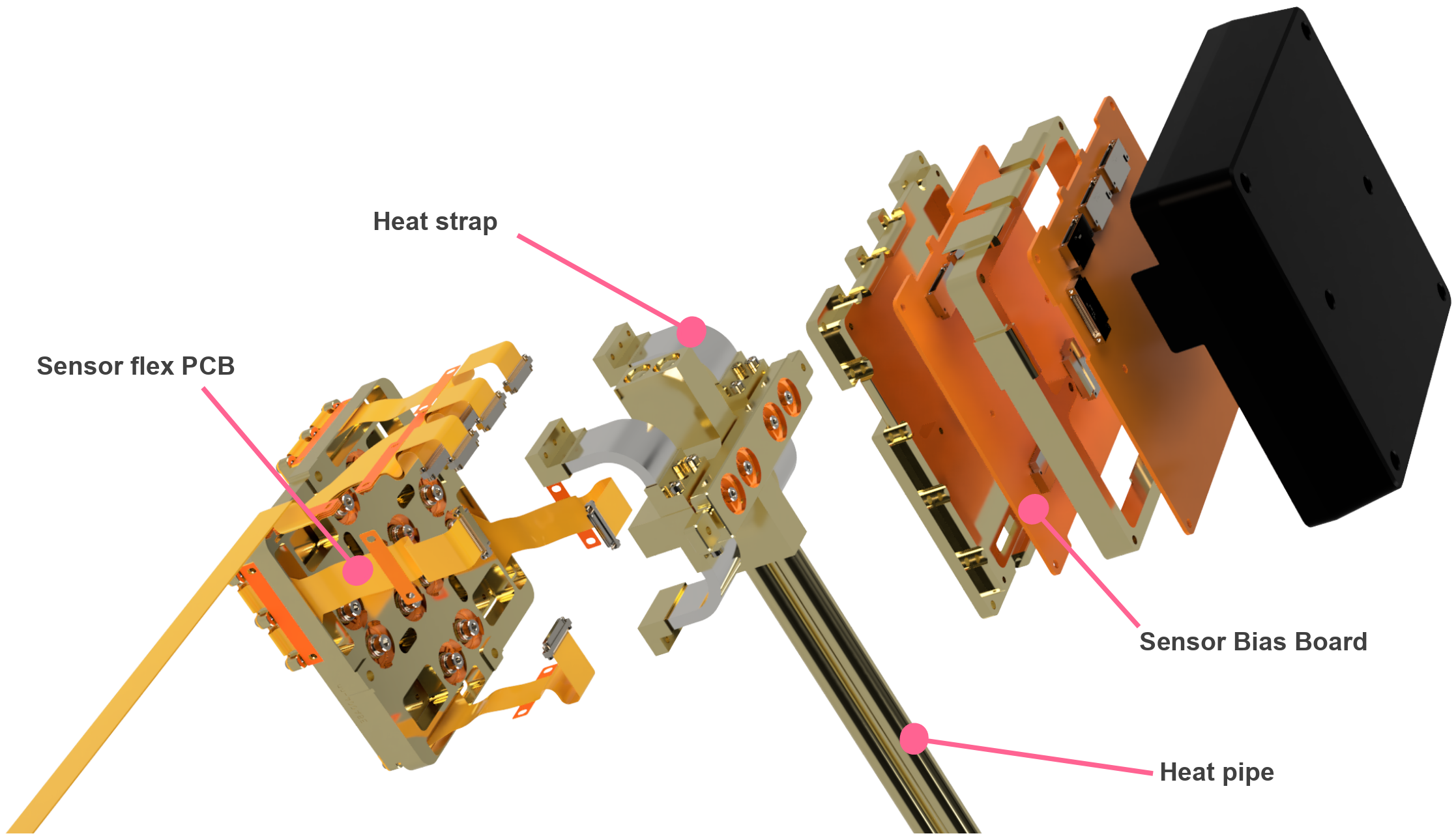}
   \end{center}
   \caption[] 
   { \label{fig:DAexploded} 
Exploded view of the \gls{DA} revealing structural, thermal and electronic components.}
   \end{figure}

The \gls{DA} structure concept is based on a milled hollow box. This ensures mechanical integrity and thus low displacements when being loaded mechanically or thermally. The hollow box will be produced of Invar 36  and provides the mechanical interfaces towards the telescope. The telescope will be held in place by four ``spiders'' attached to the sides of the DA box. Furthermore, the box provides the mechanical interface to flat-field lenses which will be placed 400~$\mu$m above the sensors. A motor will be attached at the sides of the box to allow adjustment of the lens' distance to the sensors in orbit. An outside view of the DA box is shown in Figure~\ref{fig:cam_dim}. The hollow box also provides interfaces to the interior of the \gls{DA}, where the mosaic assembly will be attached. An exploded view of the \gls{DA}'s interior is shown in Figure~\ref{fig:DAexploded}. 

Cooling of the sensor is realised by a thermal link between the sensor tiles and the thermal interface provided by the spacecraft outside of the telescope. This thermal link comprises of four graphene thermal straps (one per tile) and two ethane heat pipes. Each thermal strap is connected directly to one of the sensor tiles to mechanically decouple the mosaic from the heat pipes that transport the heat to the spacecraft’s thermal interface. The heat load transported outside of the telescope by the thermal chain is $\approx 8$~W during normal operation. The main heat loads are the radiation from the flat-field lenses ($\approx 2.8$~W), the electronic temperature control ($\approx 1.3$~W) and the \gls{CMOS} sensor periphery electronics ($\approx 1.2$~W).

To minimise heat load, each sensor is thermally decoupled by three means:  (1) the interface between sensor tiles (cold) and the mosaic plate (warm) is designed to minimise the heat conduction towards the sensor. (2) The inner surfaces of the DA box are coated to minimise the heat radiation from the DA’s structure and the mosaic plate towards the sensor tiles and the elements of the cooling chain. (3) The application of a radiation shield between the mosaic plate and the tiles to minimize the heat radiation. As mentioned, the temperature control will be performed by the spacecraft's on-board computer. To maintain the operational temperature, the use of heaters with closed loop control is foreseen. Two heaters with a maximum power of 1.5 W are allocated per sensor tile. To achieve the higher temperatures during decontamination, two additional heaters of 16~W are placed at the evaporator side of the heat pipes.

   \begin{figure} [tb]
   \begin{center}
   \begin{tabular}{c} 
   \includegraphics[width=0.50\textwidth]{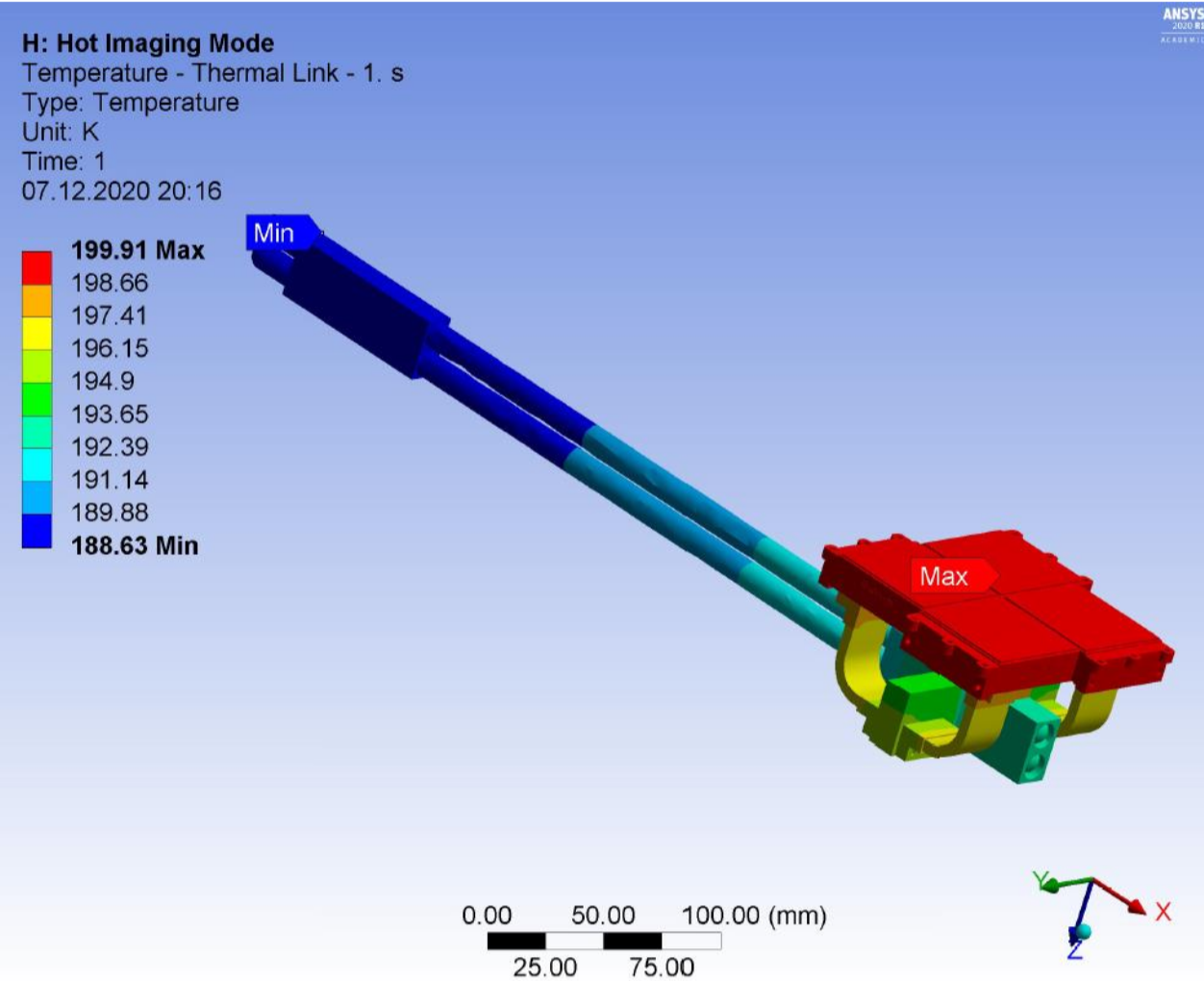}
   \includegraphics[width=0.49\textwidth]{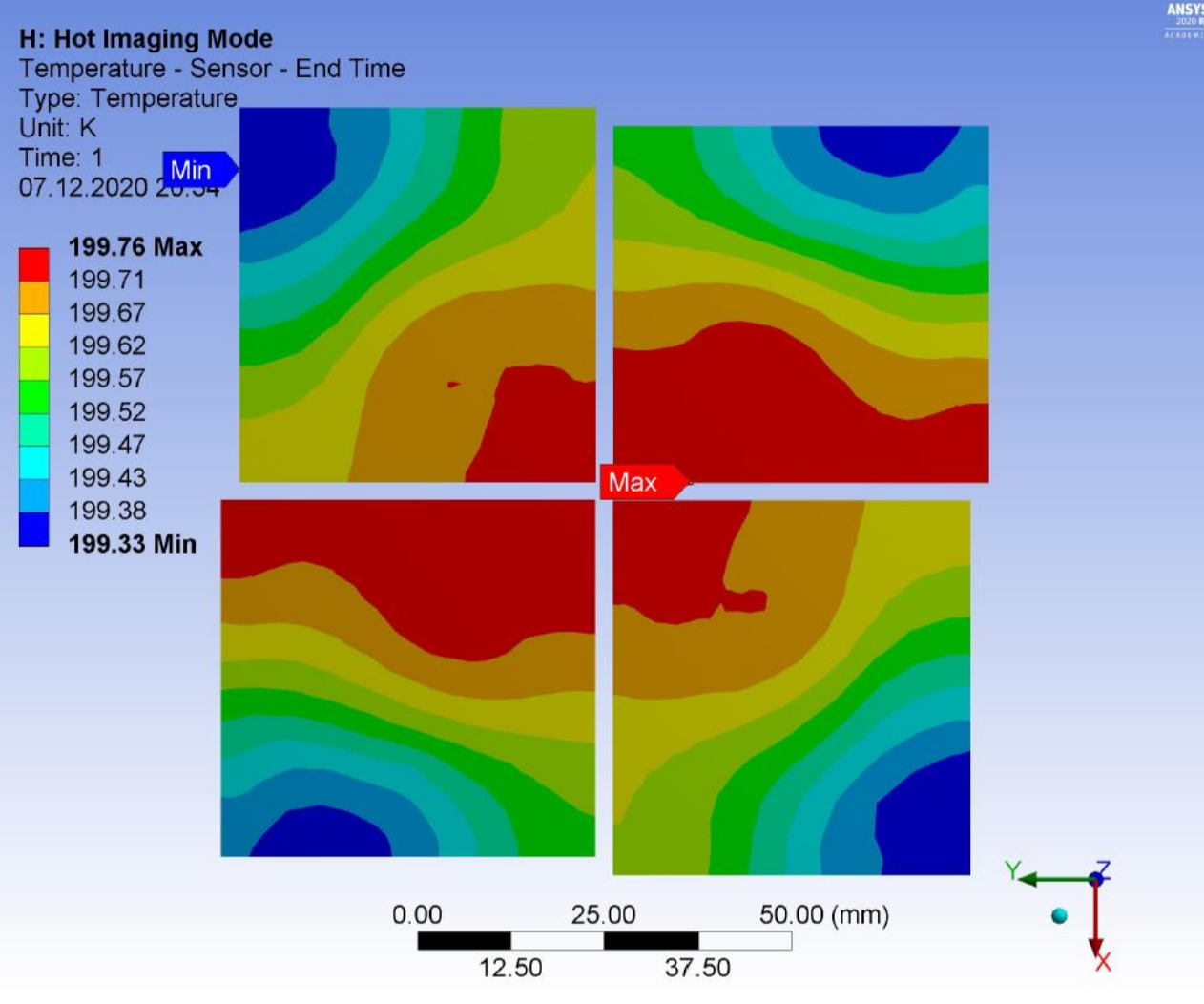}
   \end{tabular}
   \end{center}
   \caption[example] 
   { \label{fig:temp_sims} 
Thermal simulations of the ULTRASAT camera during imaging operation. The left panel shows the thermal link: the sensor tiles, heat straps, heat pipe evaporator and heat pipes. The right panel the temperature distribution on the four sensors.  }
   \end{figure} 
   
Thermal and structure simulations of the \gls{DA} were carried out using ANSYS Workbench 2020 R1. Sample simulations results are given here. The thermal simulation considered all radiative and conductive heat flows from and to the telescope, as well as to the spacecraft via the spiders. The interface between the heat pipes and the radiator is considered as a perfect heat flow sink. The simulated temperature distribution of the thermal chain and the sensors during imaging operations are shown in Figure \ref{fig:temp_sims}. The simulated temperature homogeneity of the sensors is well within the required $<$~1~K.

\begin{figure} [tb]
   \begin{center}
   \begin{tabular}{c} 
   \includegraphics[width=0.50\textwidth]{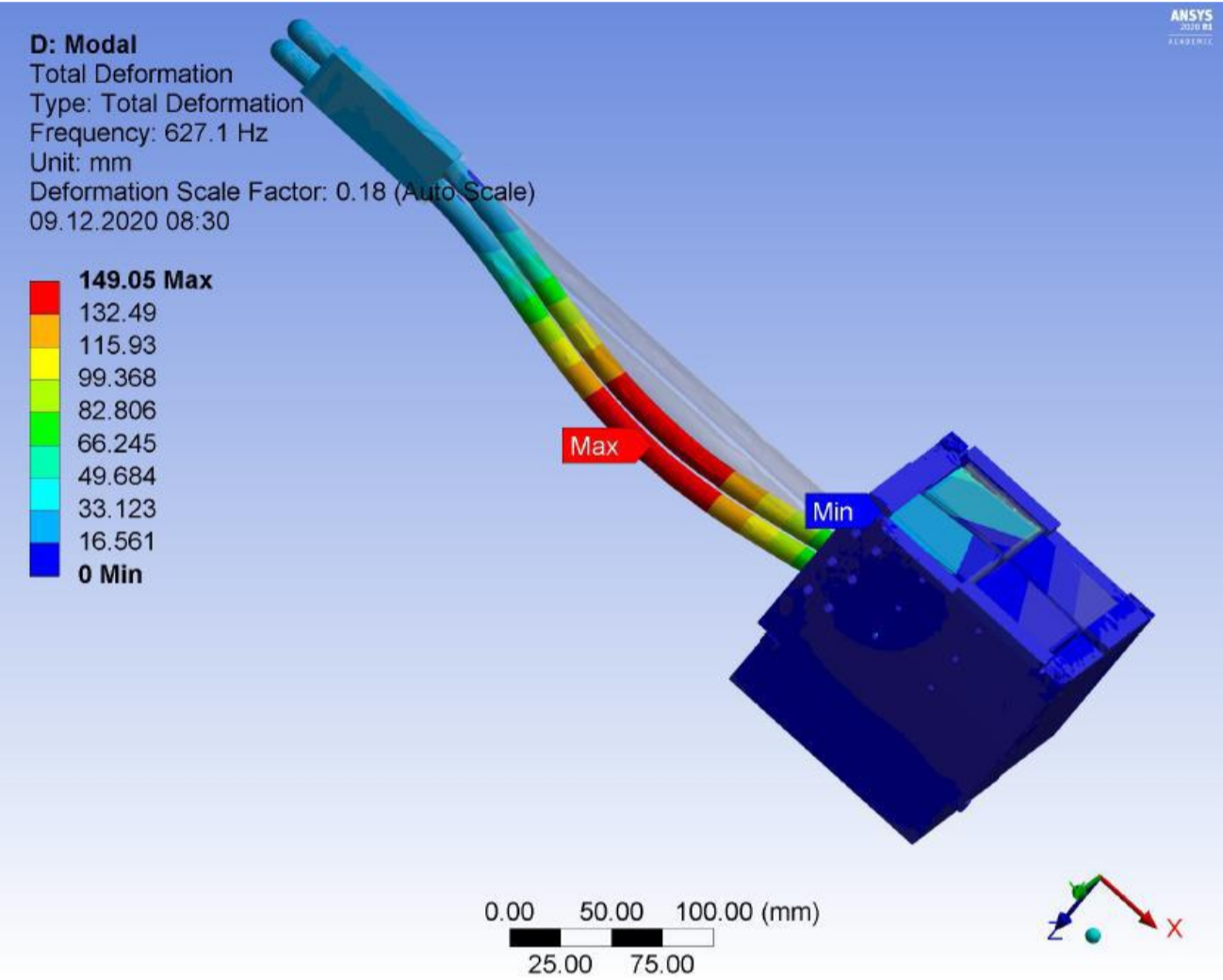}
   \includegraphics[width=0.49\textwidth]{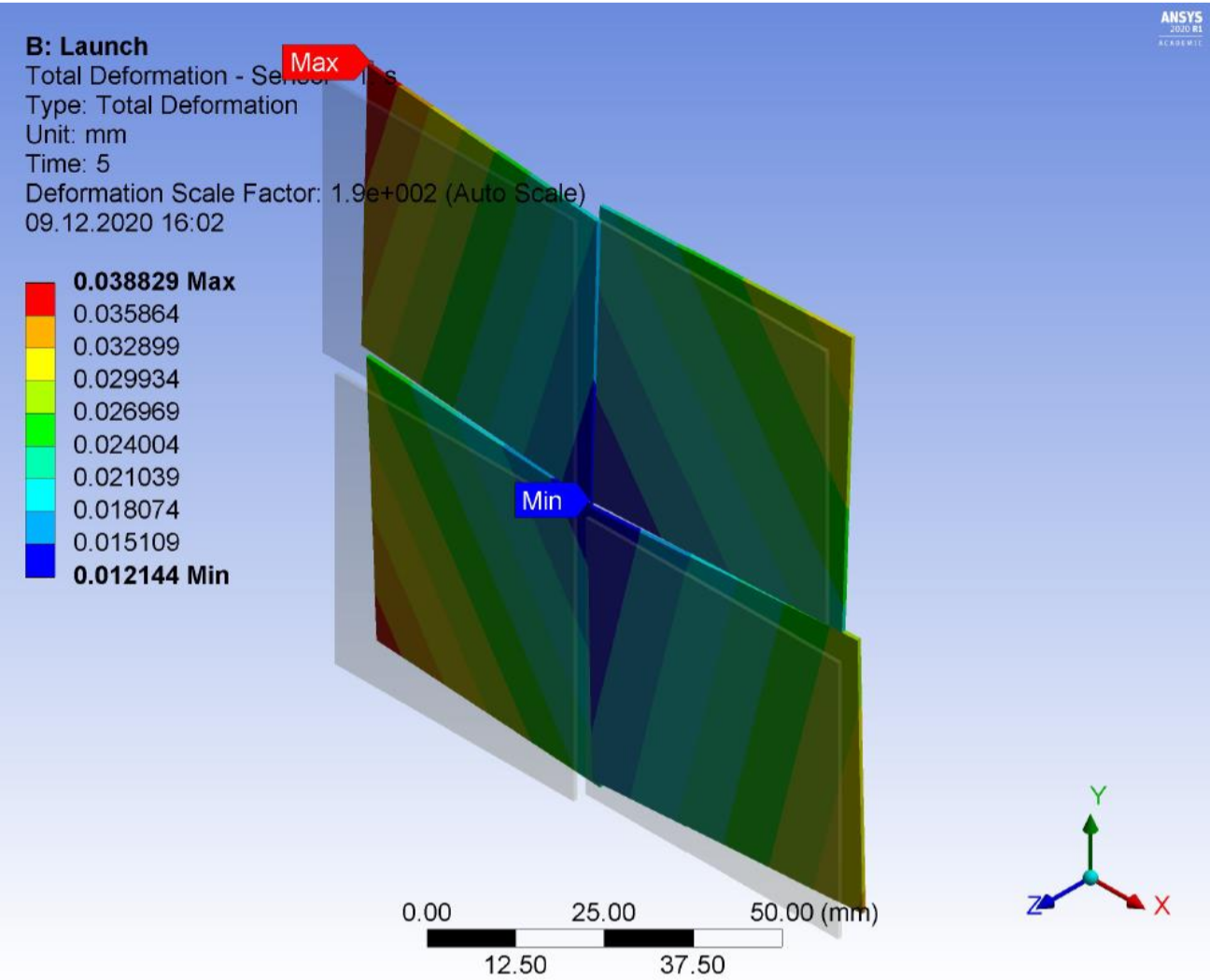}
   \end{tabular}
   \end{center}
   \caption[example] 
   { \label{fig:struct_sims} 
Two sample structural simulations of the \gls{DA}. The left panel shows the displacements at the first natural frequency of 627.1~Hz. The right panel shows the displacement distribution for 50 g acceleration in Z direction for the sensors. }
   \end{figure}

The structural simulations included static and modal analysis. The static analysis calculates the effects of steady loading conditions on the structure ignoring the inertia and damping effects, caused by time-varying loads. The load envelope required for the \gls{DA} is 50~g in each axis. All displacements were found to be within the required limits. For the camera the most critical quasi-static loads are the longitudinal and lateral g-loads during launch and the thermal strain. In particular, sensor displacements during launch has to be lower than 100~$\mu$m along the optical axis. A simulation result for this case is shown in the right panel of Figure~\ref{fig:struct_sims}. The modal analysis is performed without damping and without further boundary conditions. The analysis covers all relevant frequency bands affecting the camera during transportation and launch, from 5 up to 2,000 Hz. The first natural frequency found in the simulations is at 627.1~Hz, shown in the left panel of Figure~\ref{fig:struct_sims}. This is above the lowest permissible frequency specified in the launch requirements of 400 Hz.

\section{Readout Electronics }
\label{sec:electronics}

Each sensor is controlled and read out via two separated interfaces. The first one is an \gls{SPI}, used to load the program to the micro-controller implemented in each sensor. The program contains several parameters (such as the integration time, for example) which can be changed to optimize the detector's performance. The second one is an \gls{LVDS} interface which is used to readout the pixel data; this interface implements a one-link two-lanes, LiteFast compatible protocol. An external \gls{LVDS} clock (45 MHz nominal) needs to be provided by the \gls{FPGA}. During readout each pixel will output two 14-bits values, one for the low gain and one for the high gain. A simple threshold algorithm implemented in the \glspl{FPGA} will decide which one to keep and which one to discard.

The sensor requires seven voltages to be supplied for pixel operations plus two additional voltages for the analog and digital circuitry. The pixel supplies, in particular, have very low noise requirements ($<1$~mV RMS). Additionally, all the supplies shall be tunable in-orbit to correct for any degradation in performance caused by ageing effects. To provide all required supplies, two \glspl{SBB} are placed in close proximity ($\approx 15$~cm, see Figure~\ref{fig:DAexploded}) to the mosaic. Each \gls{SBB} includes two fully independent powering circuitry for two tiles plus monitoring for voltages and currents. The \glspl{SBB} are connected to the respective sensor tiles via a rigid-flex \gls{PCB} which also hosts  decoupling capacitors required to meet the noise requirements.

Both, the sensors and the \gls{SBB} circuitry are controlled by the \glspl{FPGA} located in the \gls{RE} box. The latter has four main functions: (1) it provides filtering and isolation of the unregulated power supply coming from the spacecraft, while lowering the voltage level (2) it configures, controls and reads out the sensor tiles (3) it controls the \gls{FME} of the flat-field lenses and (4) it receives the commands from the  spacecraft to control all the mentioned subsystems. To do so, the \gls{RE} box is connected to the spacecraft via three interfaces: (1) an unregulated DC power line (\SIrange[range-units=single]{28}{45}{\volt}) (2) a 112~kbps UART to receive  commands and send status information and (3) an link to send out the science data.

\begin{figure} [tb]
   \begin{center}
   \includegraphics[width=1\textwidth]{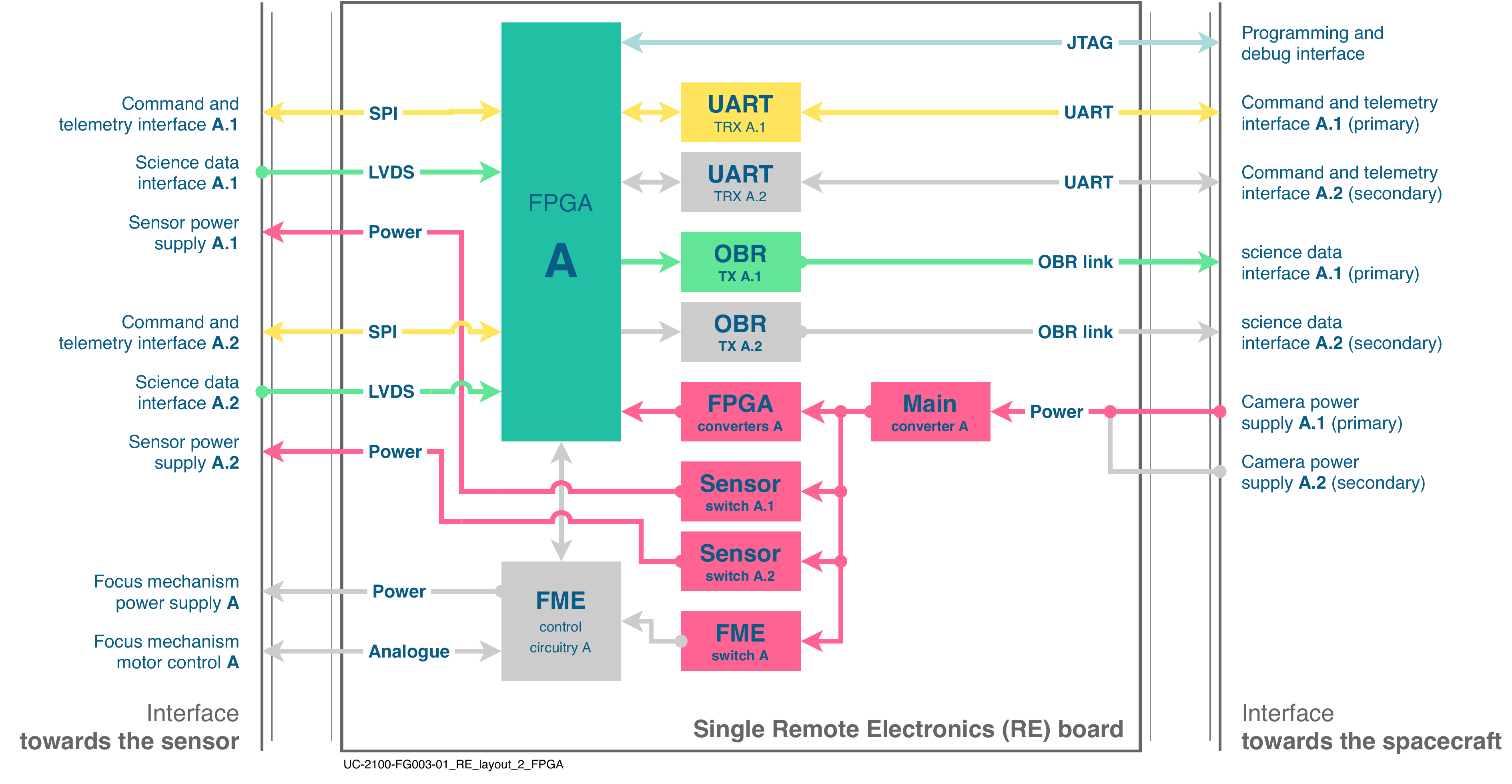}
   \end{center}
   \caption{ \label{fig:Re}
Schematic overview showing one of the two identical camera control boards that together form the \gls{RE}. The central element is the \gls{FPGA} that acts as interface element between the spacecraft and the sensor. Power is transformed by one main converter and then splits to supply the \gls{FPGA}, the associated sensor tiles and the \gls{FME} control circuitry. The schematic shows the control board in operational switching state, where the sensor is operated and the \gls{FME} is switched off. Here, all elements that are not powered during nominal sensor operation are shown in grey.}
   \end{figure}

Physically, the \gls{RE} circuitry consists of two identical \glspl{REB}, each one of them controlling and reading out half of the sensor mosaic (2 tiles) with its bias circuitry. An overview of the main interfaces and components located on one \gls{REB} is shown in Figure \ref{fig:Re}.

\section{Conclusions and outlook}
\label{sec:conclusions}

We have presented the preliminary design of the ULTRASAT camera. The design fulfills all requirements of the mission and passed the \gls{PDR} in December 2020. Since then, the design has evolved further. Two important changes have been introduced to improve the expected reliability of the camera:
\begin{enumerate}
\item The decision was taken to read out each sensor with one dedicated \gls{REB}. The number of \glspl{REB} will therefore be four. As the \glspl{SBB} are divided in four completely independent circuits this result in four fully independent camera quarters. This improves the graceful degradation of the system, as any failure in the camera electronics will only affect one sensor tile.
\item The mosaic plate will be thermally coupled to the sensor tiles. Thermal isolation is realized at the interface between the mosaic plate and the \gls{DA} box. This improves the thermo-elastic stiffness of the mosaic assembly, resulting in an improved flatness of the mosaic assembly during operation.
\end{enumerate}
The camera team is working towards the Critical Design Review, scheduled for October 2021. First camera models are expected in 2022. 

\acknowledgments 
We would like to acknowledge the help and support of the \gls{ULTRASAT} camera advisory board, composed of Andrei Cacovean, Maria F\"urmetz, Norbert Kappelmann, Olivier Limousin, Harald Michaelis, Achim Peters, Chris Tenzer, Simone del Togno, Nick Waltham and J\"orn Wilms. We would also like to express our gratitude to the Institute of Planetary Research of the DLR for their advise.

\bibliography{ultrasat_camera} 
\bibliographystyle{spiebib} 

\end{document}